\documentclass[aps,pra,showpacs,superscriptaddress,twocolumn]{revtex4}

\usepackage{mathrsfs}
\usepackage{amsfonts}
\usepackage{amssymb}
\usepackage{amsmath}
\usepackage{graphicx}
\usepackage{color}
\usepackage[colorlinks=true,citecolor=blue,linkcolor=magenta]{hyperref}
\usepackage{ulem}

\newcommand{\Tr}{\mathrm{Tr}}

\newcommand{\ket}[1]{\vert #1 \rangle}

\newcommand{\ketbra}[2]{\vert #1 \rangle \langle #2 \vert}

\newcommand{\tr}{\mathrm{Tr}}

\begin{document}

\title{Quantum computation with noisy operations}

\author{Ying Li}
%\email{ying.li.phys@gmail.com}

\affiliation{Department of Materials, University of Oxford, Parks Road, Oxford OX1 3PH, United Kingdom}

\date{\today}

\begin{abstract}
In this paper, we show how to use low-fidelity operations to control the dynamics of quantum systems. Noisy operations usually drive a system to evolve into a mixed state and damage the coherence. Sometimes frequent noisy operations result in the coherent evolution of a subsystem, and the dynamics of the subsystem is controlled by tuning noisy operations. Based on this, we find that universal quantum computation can be carried out by low-fidelity (fidelity $<90\%$) operations.
\end{abstract}

\maketitle

\section{Introduction}

Manipulating quantum systems coherently is important for quantum computation~\cite{NielsenBook}, which is believed to have non-trivial advantages over classical computation. In recent years, persistent quantum memories (e.g.~\cite{Saeedi2013}) and precise quantum operations have been demonstrated with individual qubits or clusters of a few qubits~\cite{Ladd2010,ion,SCQ}. However, quantum computation is still a challenge, and one of the main obstacles is the difficulty to maintain the fidelity of quantum operations when many qubits are assembled together. Some alternative models other than the standard model of quantum computation have been proposed to exploit different mechanisms of processing quantum information, e.g.~adiabatic quantum computation~\cite{Farhi2000}, measurement-based quantum computation~\cite{Raussendorf2001}, and dissipation quantum computation~\cite{Verstraete2009}. In this paper, we will show a protocol of quantum computation utilizing noisy operations, i.e.~the fidelity of operations is lower than $90\%$.

Decoherence and operation imperfections always induce some errors on qubits. The number of errors increases with the time, the number of operations, and the number of qubits without error correction, which could finally cause failures of quantum computing. The theory of fault-tolerant quantum computation (FTQC) predicts a threshold of the error rate: if errors occur with a rate below the threshold, errors are correctable and the computation is reliable~\cite{Knill2005}. For topological codes, the error rate threshold is about $1\%$ (one error in a hundred operations)~\cite{Raussendorf2007,Wang2011}, which is among the best records of the threshold. By combining the idea of noisy-operation quantum computation proposed in this paper and a topological code, we find that a high error rate $>10\%$ is tolerable for realistic frequency of operations and coherence time.

In our proposal of quantum computing with noisy operations, we consider a system composed of two coupled subsystems $A$ and $Q$, in which the subsystem $A$ (actuator) directly suffers noisy operations, and the subsystem $Q$ is a register storing the quantum state for processing. By frequently performing noisy operations, the subsystem $A$ is decoupled from the subsystem $Q$ and fixed in a mixed state. In this case, the subsystem $Q$ evolves solely and coherently, however, its dynamics depends on the fixed state of the subsystem $A$~\cite{OQZE}. Therefore, by changing noisy operations to alter the fixed state of the subsystem $A$, one can effectively tune the dynamics of the subsystem $Q$. Based on this new idea of the control of a quantum system, we will show that universal and scalable quantum computing can be achieved with noisy operations. By investing the fault-tolerance thresholds, we find that these operations can be very noisy.

The idea of indirect control using projective measurements~\cite{Mandilara2005}, completely-controlled dynamics~\cite{Lloyd2004,Romano2006PRA}, or initializations~\cite{Romano2006PRL} of an ancillary system has been studied theoretically and has applications in hybrid systems composed of electron spins and nuclear spins~\cite{Morton2006,Hodges2008,Mitrikas2010,Filidou2010,Zhang2011}. If those frequent operations performed on the actuator subsystem are unitary rather than noisy, both subsystems can evolve independently and coherently, which is known as the dynamical decoupling~\cite{DD}. In this paper, we first propose to use noisy operations with low fidelities rather than high-quality quantum operations as the resource for processing quantum information. In this scenario, the actuator subsystem is always fixed in a mixed state independent of the register subsystem.

\section{Quantum gates based on noisy operations}
\label{Gates}

We suppose that the free time evolution of the system is given by the Hamiltonian $H$, in which two subsystems are coupled with each other. The set of operations that can be performed on the actuator subsystem includes the initialization operation $\mathcal{I}(\cdot)=\sum_j\ketbra{0}{j}_A\cdot\ketbra{j}{0}_A$ and unitary operations $\{ \mathcal{U}(\cdot)=U\cdot U^\dag \}$. Here, $\{\ket{j}_A\}$ are the basis states, and $\{U\}$ are unitary operators of the actuator subsystem. With imperfections, operations actually performed, which are respectively denoted by $\mathcal{I}'$ and $\{\mathcal{U}'\}$ for the initialization and unitary operations, are different from those ideal operations $\mathcal{I}$ and $\{\mathcal{U}\}$. An imperfect initialization prepares the actuator in a mixed state rather than the pure state $\ket{0}_A$, and an imperfect unitary operation corrupts as well as rotates the state of the actuator subsystem.

When operations are ideal, by frequently repeating the initialization $\mathcal{I}$ followed by a unitary operation $\mathcal{U}$, the actuator subsystem is frozen in the state $\ket{\psi}_A=U\ket{0}_A$. Because only the actuator subsystem is completely frozen, the whole system evolves in a subspace with an effective Hamiltonian $H_\pi(U)=\ketbra{\psi}{\psi}_AH\ketbra{\psi}{\psi}_A$ as predicted by the quantum Zeno effect theory~\cite{Zeno,Facchi2002}. Here, the combination of the initialization and the unitary operation is similar to a projective measurement with $\ket{\psi}_A$ as the output state~\footnote{
In the limit of high frequency, both initialisation operation and projective measurement can froze the actuator system in the state $\ket{\psi}_A$. Therefore, although the initialisation operation is different from projective measurement, we still can use the theory of quantum Zeno effect to predict the effective Hamiltonian.
}. When operations are imperfect, by frequently repeating the noisy initialization $\mathcal{I}'$ followed by a noisy unitary operation $\mathcal{U}'$, the actuator subsystem is frozen in a mixed state $\rho_U=\mathcal{U}'\mathcal{I}'(\openone_A/d_A)$ rather than a pure state, where $\openone_A$ is the identity of the actuator subsystem, and $d_A$ is the dimension of the Hilbert space. Here, the combined operation $\mathcal{U}'\mathcal{I}'$ is a projector in the operator space ($\mathcal{U}'\mathcal{I}'\mathcal{U}'\mathcal{I}'=\mathcal{U}'\mathcal{I}'$), which leads to the operator quantum Zeno effect~\cite{OQZE}. Then the whole system evolves, similar to the normal quantum Zeno effect, with an effective Hamiltonian $H_\Pi(U)=\openone_A\otimes\tr_A(\rho_UH)$. In this effective Hamiltonian, two subsystems are decoupled, and the effective Hamiltonian of the register subsystem is given by $H_Q(U)=\tr_A(\rho_UH)$. Therefore, by changing the noisy unitary operation $\mathcal{U}'$, i.e.~changing the fixed state $\rho_U$ of the actuator subsystem, we can control the dynamics of the register subsystem. We would like to remark that, in order to precisely control the register subsystem, both the system Hamiltonian and noise in operations, i.e.~the map between the operation $U$ and the effective Hamiltonian $H_Q(U)$, must be known. 

In the standard model of quantum computation, a universal set of quantum gates includes a set of single-qubit gates and at least one two-qubit entangling gate such as the controlled-phase gate~\cite{NielsenBook}. In the following, we will give some examples of implementing gate operations on the register subsystem with noisy operations on the actuator subsystem. In Sec.~\ref{IandM}, we will show how to prepare and measure the register subsystem.

\subsection{Single-qubit gate}

As the first example, we consider two qubits coupled via the Heisenberg interaction $H_{\mathrm{H}}=J(\sigma_A^x\sigma_Q^x+\sigma_A^y\sigma_Q^y+\sigma_A^z\sigma_Q^z)$. These two qubits are the actuator subsystem $A$ and the register subsystem $Q$, respectively. We model imperfect operations as a combination of ideal operations and depolarizing noise. An operation with depolarizing noise reads $\mathcal{O}'=\mathcal{E}_\epsilon\mathcal{O}$, where $\mathcal{O}$ is the ideal operation, and $\mathcal{E}_\epsilon(\cdot)=(1-3\epsilon/4)\cdot+(\epsilon/4)(\sigma^x\cdot\sigma^x+\sigma^y\cdot\sigma^y+\sigma^z\cdot\sigma^z)$ is the depolarizing operation. In this model of noise, the ideal operation is performed with the probability $1-\epsilon$, while the state of the qubit is completely destroyed, i.e.~turns into the maximally mixed state, with the probability $\epsilon$. Under frequent noise operations, the fixed state of the actuator qubit can always be written as $\rho_U=(1/2)\openone_A+p_x\sigma_A^x+p_y\sigma_A^y+p_z\sigma_A^z$. Then, the corresponding effective Hamiltonian of the register qubit is $H_Q(U)=J(p_x\sigma_Q^x+p_y\sigma_Q^y+p_z\sigma_Q^z)$. If only the initialization is performed, the fixed state is $\rho_{\openone}=(1-\epsilon_i/2)\ketbra{0}{0}_A+\epsilon_i/2\ketbra{1}{1}_A$, and the corresponding effective Hamiltonian is $H_Q(\openone)=(1-\epsilon_i)J\sigma_Q^z$, where $\epsilon_i$ is the depolarizing rate of the noisy initialization. By combining the initialization with a Hadamard gate, the fixed state is changed to $\rho_{H}=(2-\epsilon_i-\epsilon_h+\epsilon_i\epsilon_h)/2\ketbra{+}{+}_A+(\epsilon_i+\epsilon_h-\epsilon_i\epsilon_h)/2\ketbra{-}{-}_A$, and the corresponding effective Hamiltonian is $H_Q(H)=(1-\epsilon_i)(1-\epsilon_h)J\sigma_Q^x$. Here, $\ket{\pm}_A=(1/\sqrt{2})(\ket{0}_A\pm\ket{1}_A)$, and $\epsilon_h$ is the depolarizing rate of the noisy Hadamard gate. Therefore, in this example, the initialization and Hadamard gate on the actuator qubit are enough for universal single-qubit gates on the register qubit.

\begin{figure}[tbp]
\includegraphics[width=1\linewidth]{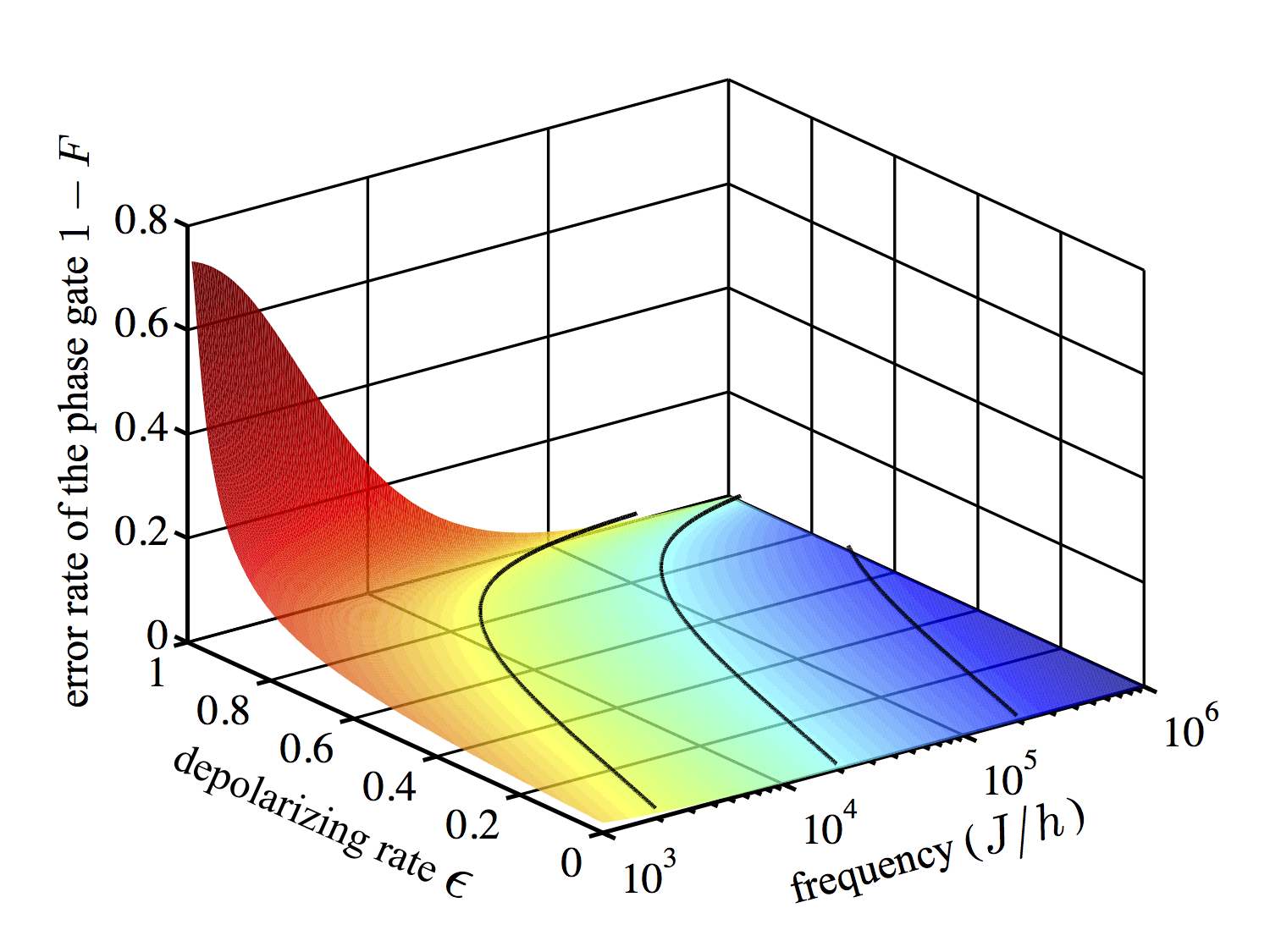}
\caption{
Error rate of a single-qubit phase gate $\sigma^z$ on the register qubit implemented with frequent noisy operations on the actuator qubit. $F$ is the entanglement fidelity~\cite{Nielsen2002} of the phase gate. The error rate ($1-F$) always increases with the depolarizing rate and decreases with the frequency of noisy operations. Three contour lines correspond to phase-gate fidelities $99\%$, $99.9\%$, and $99.99\%$, respectively. See Sec.~\ref{AppTS} for details.
}
\label{FidErr}
\end{figure}

When the frequency of noisy operations is finite, the evolution of the register subsystem is not exactly described by the effective Hamiltonian, which may result in errors in the register subsystem. In the previous example of implementing single-qubit gates, a phase gate $\sigma_Q^z$ on the register qubit is equivalent to the time evolution described by $\exp[-iH_Q(\openone)t/\hbar]$ with $t=\hbar\pi/[2(1-\epsilon_i)J]$. When the frequency of noisy operations is finite, the fidelity of the phase gate decreases with the depolarizing rate and increases with the frequency of noisy operations as shown in Fig.~\ref{FidErr}. We find that even if the depolarizing rate is very high (e.g.~$80\%$), a high-fidelity (e.g.~$99\%$) phase gate can still be achieved (e.g.~with the frequency $\sim 1.5\times 10^4 J/h$). Actually, when the frequency approaches infinite, the depolarizing rate only affects the time cost of implementing the phase gate, i.e.~is limited by the coherence time.

\subsection{Two-qubit gate}

For the two-qubit entangling gate, we consider three qubits coupled via a three-qubit Ising interaction $H_{\mathrm{I}}=J\sigma_A^z\sigma_{Q1}^z\sigma_{Q2}^z$, where qubits $Q1$ and $Q2$ form the register subsystem, and qubit $A$ forms the actuator subsystem. With frequently initializing the actuator qubit, the dynamics of two register qubits is given by the effective two-qubit Ising interaction $H_{QQ}(\openone)=J(1-\epsilon_i)\sigma_{Q1}^z\sigma_{Q2}^z$. And the time evolution $\exp[-iH_{QQ}(\openone)t/\hbar]$ with $t=\hbar\pi/[4(1-\epsilon_i)J]$ gives the two-qubit phase gate $R_{ZZ}=(\openone-i\sigma_{Q1}^z\sigma_{Q2}^z)/\sqrt{2}$, which can maximally entangle two register qubits and is identical to the standard controlled-phase gate up to single-qubit phase gates. In Sec.~\ref{twobody}, we will show how to implement a two-qubit gate with only two-body interactions.

\begin{figure}[tbp]
\includegraphics[width=1\linewidth]{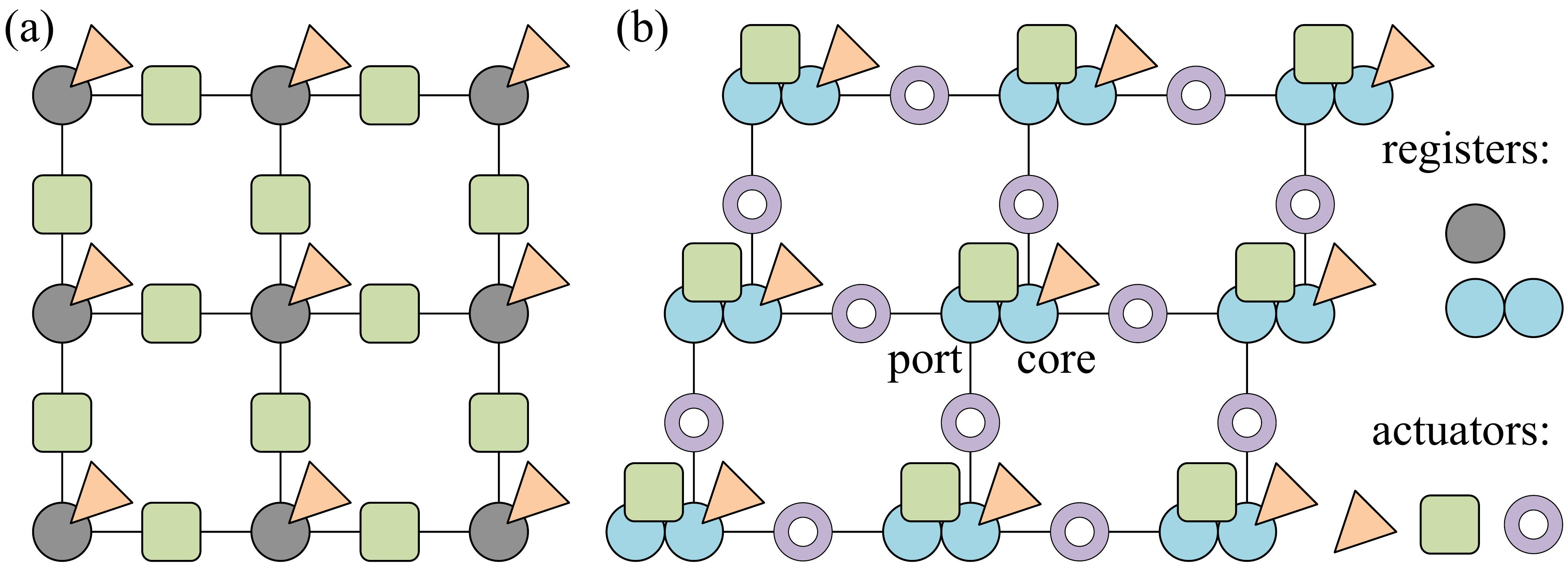}
\caption{
Networks for universal quantum computation with noisy operations. (a) A network that allows universal quantum computation. Each register qubit (round) is coupled with one actuator qubit (triangle) for controlling single-qubit gates and four actuator qubits (squares) for controlling two-qubit gates. (b) A network with only two-body interactions. Each double-round register represents a four-state system encoding two qubits. The triangle actuator is coupled with only one qubit of the register (the core qubit), and the square actuator is coupled with both qubits (the core qubit and the port qubit). Double-round registers are connected by two-body interactions with actuator qubits (rings) for controlling the state transfer.
}
\label{network}
\end{figure}

\section{Universal quantum computation with noisy operations}

To have a universal quantum computer, we need to integrate single-qubit gates and two-qubit gates in the same scalable network. For example, in the two-dimensional network shown in Fig.~\ref{network} (a), each register qubit (round) is coupled with an actuator qubit (triangle) via the Heisenberg interaction $H_{\mathrm{H}}$ for single-qubit gates, and each pair of neighbouring register qubits are coupled with an additional actuator qubit (square) via the three-qubit Ising interaction $H_{\mathrm{I}}$ for two-qubit gates. In this network, all register qubits form the register subsystem, and all actuator qubits form the actuator subsystem. We have discussed how to implement single-qubit gates and the two-qubit phase gate via the Heisenberg interaction and the three-qubit Ising interaction, respectively. Gate operations on register qubits are switched off (switched to $\openone$) by frequently performing the single-qubit twirling operation $\mathcal{E}_1$ (a depolarizing operation with the rate $1$) on corresponding actuator qubits. The twirling operation is equivalent to randomly performing Pauli gates. With frequent twirling operations, an actuator qubit is frozen in the maximally-mixed state $\rho=(\ketbra{0}{0}_A+\ketbra{1}{1}_A)/2$, and then the corresponding effective dynamics is switched off (the effective Hamiltonian $\propto \openone$). By changing between twirling operations and other noisy operations, i.e.~initialization and unitary-gate operations, a quantum circuit can be implemented in this network.

\subsection{Two-body-interaction model}
\label{twobody}

\begin{figure}[tbp]
\includegraphics[width=1\linewidth]{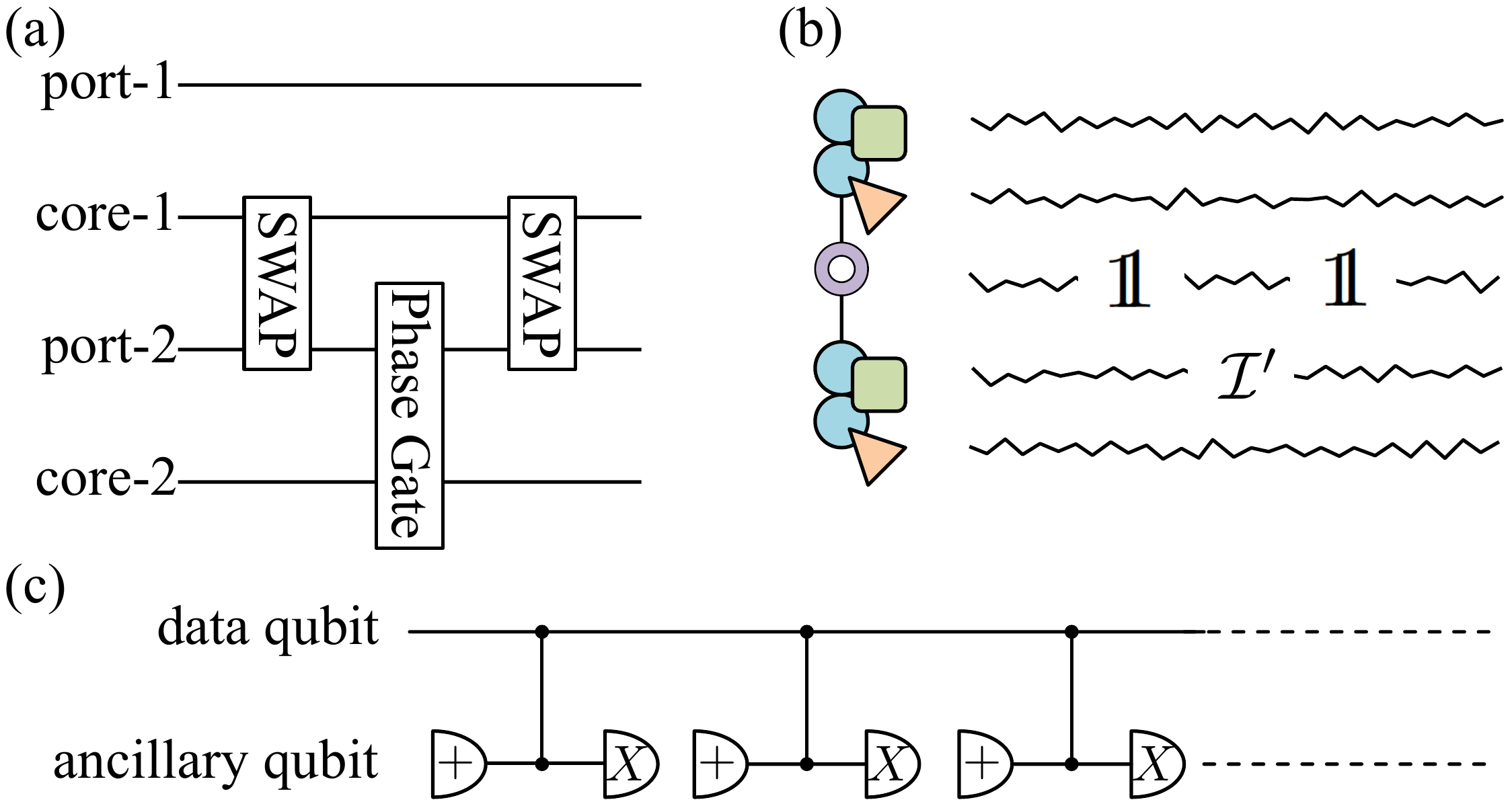}
\caption{
Circuits for (a) the effective two-qubit phase gate on neighbouring complexes, (b) the corresponding operations on actuator qubits, and (c) the measurement distillation. In (b), zigzag lines represent frequent twirling operations $\mathcal{E}_1$, $\openone$ denotes turning off the twirling operation, and $\mathcal{I}'$ denotes frequent initialization operations. In (c), both the data qubit and the ancillary qubit are register qubits, and controlled phase gates are implemented with frequent noisy operations on the actuator subsystem. In each round of the distillation, the ancillary qubit is initialized in the state $\ket{+}$ and measured in the $\sigma^x$ basis.
}
\label{circuit}
\end{figure}

The network for universal quantum computation can also be built with only two-body interactions. In the network shown in Fig.~\ref{network} (b), the elementary unit is a complex composed of two actuator qubits (triangle and square) and a four-state register particle (double round), e.g.~a spin-$3/2$ particle. Here two qubits are encoded in each register particle, which are respectively called the core qubit and the port qubit. The triangle actuator qubit is coupled to the core qubit via $H_{\mathrm{H}}$. And the square actuator qubit is coupled to both qubits via $H_{\mathrm{I}}$, which, however, is a two-body interaction because two qubits are in the same particle. Register particles are connected via the third kind of actuator qubits (rings). The coupling between a pair of neighbouring register particles and a ring actuator qubit is the XY-interaction $H_{\mathrm{XY}}=J[\sigma_A^x(\sigma_{C1}^x+\sigma_{P2}^x)+\sigma_A^y(\sigma_{C1}^y+\sigma_{P2}^y)]$, where complex-$1$ and complex-$2$ are two neighbouring complexes, $C1$ and $P2$ are respectively the core qubit of complex-$1$ and the port qubit of complex-$2$, and $A$ is the ring actuator qubit. In this network, the information processed in the quantum computing is stored in core qubits. Single-qubit gates on core qubits are implemented via triangle actuator qubits. Two-qubit gates are achieved by the state transfer between core qubits and port qubits in neighbouring complexes. The XY-interaction $H_{\mathrm{XY}}$ drives a time evolution ending up with a swap gate between two qubits $C1$ and $P2$ at the time $t=\hbar\pi/2\sqrt{2}J$. By frequently performing the twirling operation $\mathcal{E}_{1}$ on the ring actuator qubit, the XY-interaction can be effectively switched off, and complexes are decoupled. When a two-qubit phase gate on core qubits $C1$ and $C2$ is required, the twirling operation is turned off for the time $t$. Then the free time evolution transfers the state from the core qubit $C1$ to the port qubit $P2$. After a local two-qubit phase gate on qubits $C2$ and $P2$ (via the square actuator qubit) and another swap gate, a two-qubit phase gate on $C1$ and $C2$ is achieved [see the circuit in Figs.~\ref{circuit}(a)~and~\ref{circuit}(b)]. We would like to remark that, each gate given by the time evolution
$$
e^{-\frac{i}{\hbar}H_{\mathrm{XY}}t}=\frac{1}{2}(\sigma_{A}^z\sigma_{C1}^z+\sigma_{A}^z\sigma_{P2}^z+\sigma_{C1}^z\sigma_{P2}^z-\openone)\mathrm{SWAP}
$$
has an additional phase to the net swap gate $\mathrm{SWAP}=(\sigma_{C1}^x\sigma_{P2}^x+\sigma_{C1}^y\sigma_{P2}^y+\sigma_{C1}^z\sigma_{P2}^z+\openone)/2$ depending on initial states of the port qubit $P2$ and the ring actuator qubit $A$. Fortunately, the additional phases attached with two swap gates are cancelled with each other as they commute with the two-qubit phase gate. Therefore, the overall operation on two core qubits $C1$ and $C2$ is independent of the initial states of the port qubit $P2$ and the ring actuator qubit $A$.

\subsection{Initialisation and measurement of register qubits}
\label{IandM}

Besides gate operations, preparation and readout of the state of qubits are also required by quantum computing. Here we suppose that register qubits can be directly initialized and measured but with a low fidelity. If gate operations on register qubits can be implemented with high fidelity, we can measure the state of a register qubit precisely with the help of a distillation circuit [see Fig.~\ref{circuit}~(c)]. In the ideal situation that there is no error in gate operations on register qubits, an accurate effective measurement on the data qubit can always be achieved by repeating the circuit for many times, and the state of the data qubit can be read (in the $\sigma^z$ basis) from the majority of measurement outcomes of the ancillary qubit. When gate operations also have errors, the protocol of reading the data qubit can be adapted to improve its fidelity (see Sec.~\ref{AppTS}). Once the state of the data qubit is successfully measured, the data qubit is initialized in the state either $\ket{0}$ or $\ket{1}$.

\begin{figure}[tbp]
\includegraphics[width=0.9\linewidth]{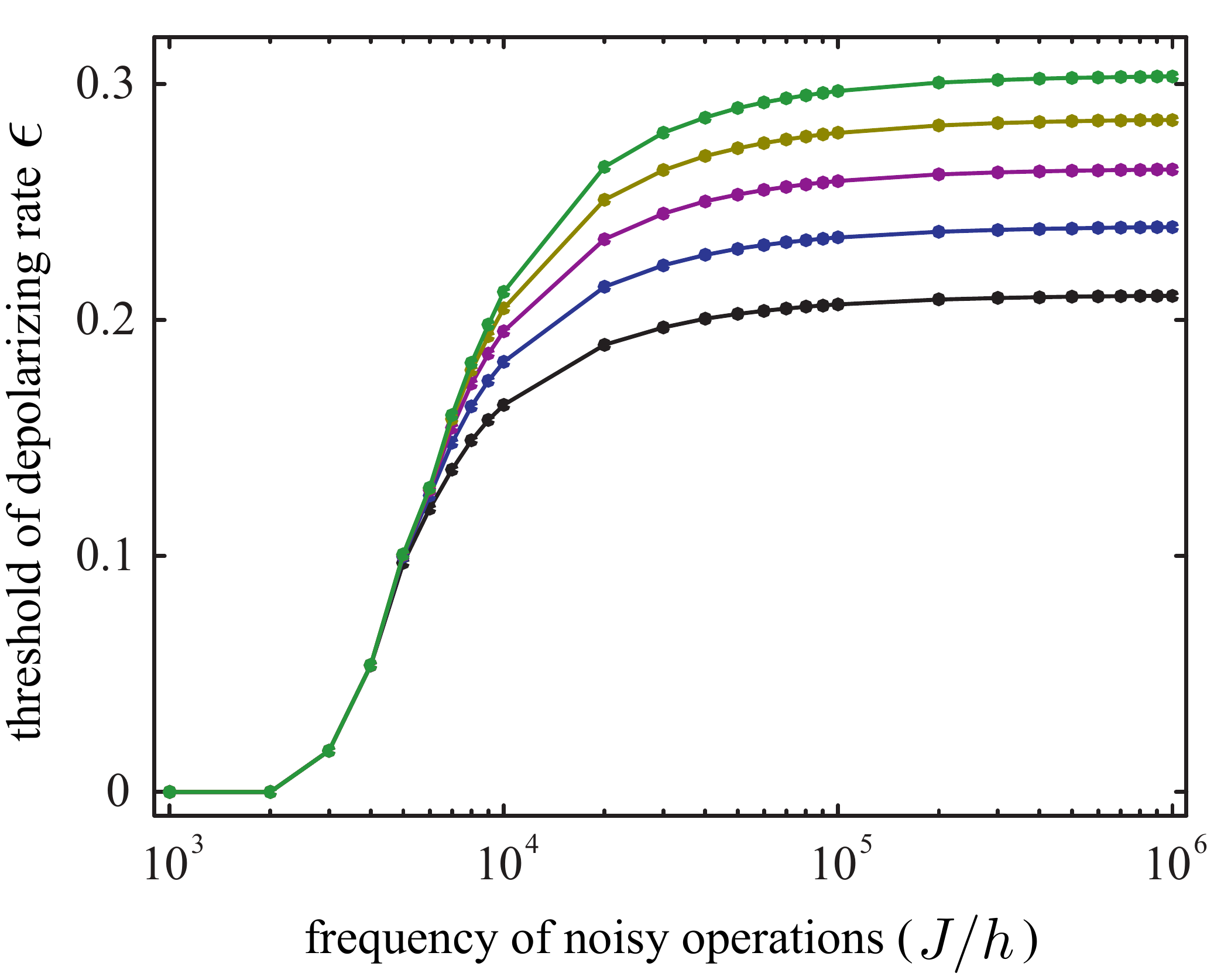}
\caption{
Thresholds of the depolarizing rate in noisy-operation quantum computation. From bottom to top the lines correspond to distilling measurements for $9$, $11$, $13$, $15$, and $17$ rounds, respectively. Below the threshold, errors on register qubits can be corrected with the error correction code. We would like to remark that thresholds could be further improved by considering higher frequencies and more rounds of distillations. To obtain these thresholds, we have assumed that operations performed on actuator qubits and register qubits are all noisy, and noises are depolarized with the same depolarizing rate.
}
\label{threshold}
\end{figure}

\section{Fault-tolerant quantum computation and thresholds}

Finally, we would like to show FTQC thresholds of noisy operations in Fig.~\ref{threshold}. For these thresholds, we have considered using a three-dimensional version of the network shown in Fig.~\ref{network}~(a) to generate a topology-protected cluster state (see Sec.~\ref{AppTS} for details), in which errors can be corrected if less than $3\%$ of qubits are affected by errors~\cite{RaussendorfAP}. We also have assumed that all operations are noisy, and noises are depolarized with the same depolarizing rate. In the result, for a frequency of $10^4 J/h$, the depolarizing rate threshold is about $20.1\%$, which corresponds to the error rate $10.05\%$ for the noisy initialization and measurement and the error rate $15.075\%$ for noisy unitary operations. To obtain the thresholds, we have neglected the environment-induced decoherence. In the $17$-round distillation case, all operations on register qubits from initialization to measurement are finished within the time $7(J/h)^{-1}$. Hence, if the coherence time is much longer than $7(J/h)^{-1} / 3\% \simeq 233(J/h)^{-1}$, errors induced by decoherence occur with a rate much lower than $3\%$ and only reduce the thresholds slightly. We have also neglected the fluctuation of interactions and operation noises. A small fluctuation ($\ll 1\%$) of the interaction strength $J$ and the noise parameter $\epsilon$ will not affect the threshold significantly.

\section{Discussions and summary}

A candidate for realizing noisy-operation quantum computation is the kind of hybrid systems composed of electron spins and nuclear spins~\cite{Morton2006,Hodges2008,Mitrikas2010,Filidou2010,Zhang2011}. Electron spins in quantum dots can be initialized and manipulated in tens of picoseconds~\cite{Godden2010,Berezovsky2008} and can play the role of actuators. Nuclear spins are usually well decoupled from the microwave and optical pumping for electron-spin operations and can have an extremely long coherence time (e.g.~hours~\cite{Saeedi2013}), hence, can play the role of registers. With the coupling strength of $1\text{ MHz}$ between electron spins and nuclear spins, provided noisy operations on electron spins can be performed in $100\text{ ps}$, one can operate nuclear spins in $\sim 1\text{ }\mu\text{s}$.

In summary, we have discussed how to use noisy operations to process quantum information and obtained FTQC thresholds in an example model. We find that fidelities of operations can be even lower than $90\%$. Our results provide a way to achieve quantum computation by boosting the operation frequency rather than the operation fidelity. However, we would like to remark that this protocol of noisy-operation quantum computation cannot replace error correction codes. Actually, the error correction is an important component of the overall protocol. Because quantum computation is the most complicated task among other applications of quantum technologies, we believe that the same idea can also be used in quantum communication and quantum sensing. In this paper, we have focused on isotropic Heisenberg interactions and Ising-type interactions as examples. The same idea can be applied to other types of interactions, e.g.~interactions between electron spins and nuclear spins could be anisotropic. Although we have only discussed depolarising noise in detail, noises are not restricted to a specific type. Necessary conditions for noises are given in Sec.~\ref{AppNC}, and only a small set of noises, in which operations are extremely noisy, are not suitable for noisy-operation quantum computation.

\begin{acknowledgments}
This work was supported by the EPSRC platform grant `Molecular Quantum Devices' (EP/J015067/1). I would like to thank Simon C. Benjamin for helpful discussions.
\end{acknowledgments}

\appendix

\section{Threshold simulation}
\label{AppTS}

\subsection{Model}

\begin{figure}[tbp]
\includegraphics[width=8.0 cm]{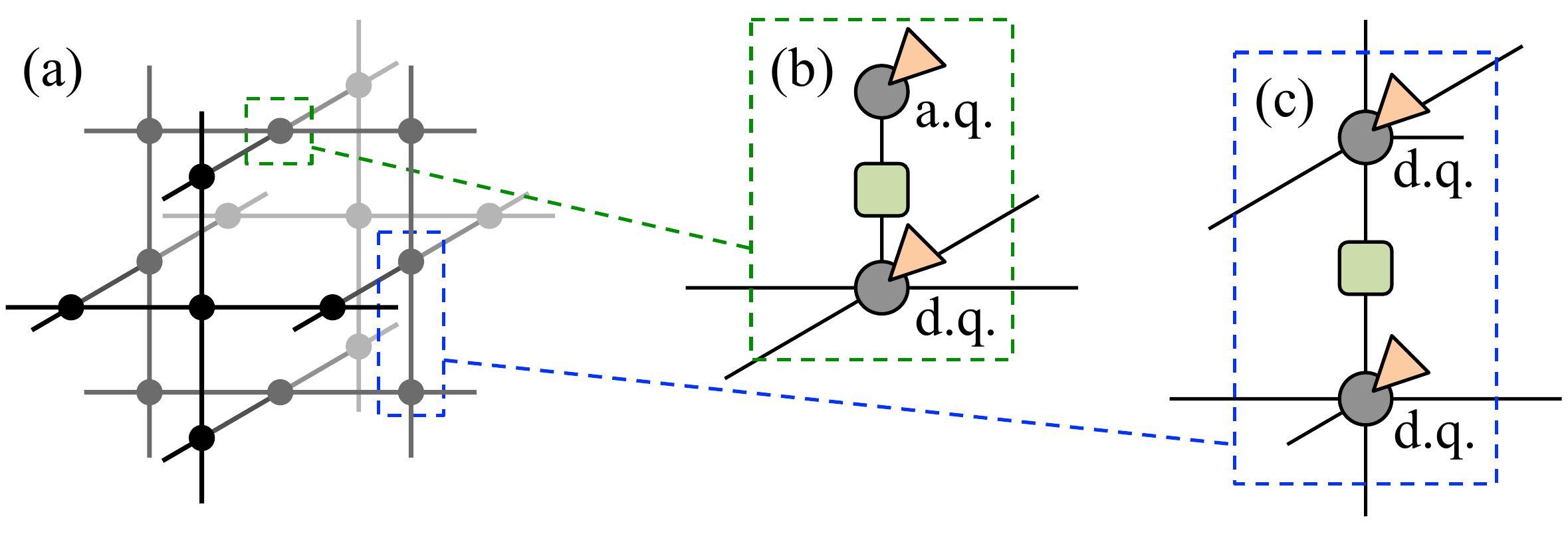}
\caption{
A three-dimensional NOQC network for building the topology-protected cluster state. (a) An elementary cubic of the topology-protected cluster state. On the cluster state, each vertex represents a qubit initialized in the state $\ket{+}$, and each edge represents a controlled-phase gate on corresponding qubits. (b) Each qubit on the cluster state corresponds to a data qubit (d.q.) on the NOQC network. An ancillary qubit (a.q.) is associated with every data qubit for initialization and measurement distillations. (c) Neighbouring data qubits are connected according to the cluster state lattice. Here, both data qubits and ancillary qubits are register qubits, which are controlled by noisy operations applied on actuator qubits.
}
\label{3Dnetwork}
\end{figure}

To obtain the thresholds of fault-tolerant quantum computation (FTQC) on the noisy-operation quantum computation (NOQC) architecture, we consider building the topology-protected cluster state [see Fig.~\ref{3Dnetwork} (a)], which can tolerate phase errors with a rate of $3\%$~\cite{RaussendorfAP}. To build the topology-protected cluster state, we need a three-dimensional version of the network shown in Fig.~\ref{network}~(a). On the three-dimensional network (see Fig.~\ref{3Dnetwork}), each cluster-state qubit corresponds to a data register qubit, and an ancillary register qubit is attached to every data qubit for initialization and measurement distillations. Each register qubit is coupled with a triangle actuator qubit for single-qubit gates, and each pair of connected register qubits are coupled with a square actuator qubit for two-qubit gates. The interactions between qubits, $H_{\mathrm{H}}$ and $H_{\mathrm{I}}$, are discussed in Sec.~\ref{Gates}. We suppose that, the initialization and measurement in the $\sigma^z$ basis $\{\ket{0},\ket{1}\}$, the Hadamard gate, and Pauli gates could be applied on actuator qubits, the initialization and measurement in the $\sigma^x$ basis $\{\ket{+},\ket{-}\}$ could be applied on register qubits, and all of these operations have the same depolarizing rate $\epsilon$ and could be repeated with the same frequency $f$. Here, a noisy measurement is modelled as a depolarizing error followed by a perfect measurement.

\subsection{Circuit}

\begin{figure*}[tbp]
\includegraphics[width=14.0 cm]{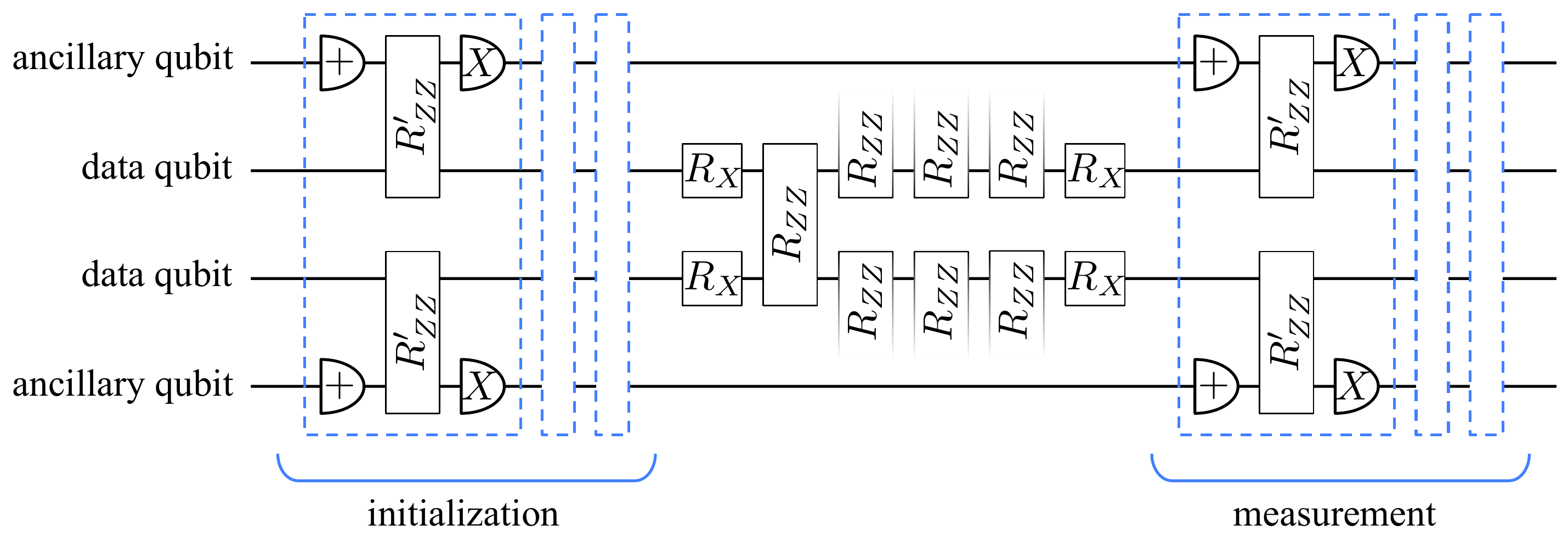}
\caption{
Overall circuit for the preparation and measurement of a cluster state on the NOQC network. In this circuit, the two cluster-state qubits are effectively measured in the $\sigma^x$ basis, which is the most likely case in the measurement-based quantum computation on the topology-protected cluster state.
}
\label{CScircuit}
\end{figure*}

The cluster state is prepared and measured with the overall circuit shown in Fig.~\ref{CScircuit}. First, data qubits are initialized in the $\sigma^z$ basis by repeating the distillation circuit. After initialization, data qubits are in states either $\ket{0}$ or $\ket{1}$ depending on measurement outcomes in distillation circuits. Flip operations could be used to align all data qubits to the state $\ket{0}$, but they are not necessary because one can update the basis rather than physically implement flip operations. Second, data qubits are further prepared in the state $\ket{+}$ via single-qubit gates, and then a cluster state can be built on the network with two-qubit phase gates. Finally, data qubits are measured in the $\sigma^z$ basis by another set of distillation circuits. In Fig.~\ref{CScircuit}, qubits on the prepared cluster state are effectively measured in the $\sigma^x$ basis because of single-qubit gates between two-qubit phase gates and measurement distillation circuits.

For each round of the distillation (each blue box in Fig.~\ref{CScircuit}), ancillary qubits are initialized in the state $\ket{+}$ and measured in the $\sigma^x$ basis. The two-qubit gate $R_{ZZ}'=R_{Z_{\mathrm{a.q.}}}R_{ZZ}=(\openone-\sigma^z_{\mathrm{d.q.}})/2-i(\openone+\sigma^z_{\mathrm{d.q.}})\sigma^z_{\mathrm{a.q.}}/2$ is a phase gate on the ancillary qubit depending on the state of the data qubit and equivalent to the standard controlled-phase gate upto a single-qubit phase gate on the data qubit. Here, $\sigma^z_{\mathrm{d.q.}}$ and $\sigma^z_{\mathrm{a.q.}}$ are Pauli operators of the data qubit and the ancillary qubit, respectively, $R_{Z}=(\openone-i\sigma^z)/\sqrt{2}$ is a single-qubit phase gate, which is implemented by frequently performing noisy initializations on the corresponding triangle actuator, and the two-qubit phase gate $R_{ZZ}$ is implemented by frequently performing noisy initializations on the corresponding square actuator. As $R_{Z_{\mathrm{a.q.}}}$ and $R_{ZZ}$ commute with each other, the two gates are implemented at the same time. Each distillation circuit is effectively a measurement of the data qubit in the $\sigma^z$ basis, hence by repeating the circuit, data qubits are initialized or measured in the $\sigma^z$ basis.

The single-qubit gate $R_ZR_X$ could be used to rotate the data-qubit state from $\ket{0}$ ($\ket{1}$) to $\ket{+}$ ($\ket{-}$) for the cluster-state generation. And the single-qubit gate $R_XR_Z^\dag$ could be used to rotate the data-qubit state from $\ket{+}$ ($\ket{-}$) to $\ket{1}$ ($\ket{0}$) for effective measurements in the $\sigma^x$ basis. Here, the gate $R_{X}=(\openone-i\sigma^x)/\sqrt{2}$ is implemented by frequently and alternatively performing noisy initializations and Hadamard gates on the corresponding triangle actuator. In the circuit shown in Fig.~\ref{CScircuit}, $R_Z$ and $R_Z^\dag$ are cancelled by each other as they commute with the two-qubit phase gate $R_{ZZ}$. The standard controlled-phase gate $\Lambda_{Z}=(\openone-\sigma^z_1)/2+(\openone+\sigma^z_1)\sigma^z_2/2$ for the cluster-state generation is replaced by the two-qubit phase gate $R_{ZZ}=(\openone-i\sigma^z_1\sigma^z_2)/\sqrt{2}$. There are a total of four controlled-phase gates applied on each data qubit, which are implemented at the same time as they commute with each other. Because $\Lambda_{Z}=R_{Z_1}^\dag R_{Z_2}^\dag R_{ZZ}$, the four single-qubit phase gates are cancelled as $R_{Z}^{\dag 4}=\openone$, and each controlled-phase gate $\Lambda_{Z}$ is replaced by a two-qubit phase gate $R_{ZZ}$.

\subsection{Errors}

For preparing the cluster state and measuring cluster-state qubits in the $\sigma^x$ basis, the noisy-operation-controlled operations (NOCOs) implemented on register qubits includes $R_{ZZ}'$, $R_X$, $R_{ZZ}$, and decoupling operations. Because we are interested in the case that the rate of errors on each cluster-state qubit is $\sim 3\%$, the error rate of each individual NOCO is $\lesssim 3\%$. Therefore, the probability of two errors occurring on the same cluster-state qubit but induced by different operations is $\lesssim 0.09\% \ll 3\%$. In the numerical simulations of error rates, we will neglect the possibility of the case that two errors occur on the same cluster-state qubit, i.e.~when we study the errors induced by one NOCO, we suppose all other NOCOs are perfectly performed.

For a NOCO $R$ implemented with the noisy operation $\mathcal{U}'\mathcal{I}'$ and the interaction $H$, the operation actually performed on register qubits reads $\mathcal{R}'(\cdot) = \Tr_A [(\mathcal{T}\mathcal{U}'\mathcal{I}')^N(\cdot \otimes \rho_A )]$, where $\mathcal{T}(\cdot)=e^{-iH/(f\hbar)}\cdot e^{iH/(f\hbar)}$ denotes the free time evolution, and $f$ is the frequency of repeating noisy operations. Here, we have supposed that the time of performing a noisy operation is much shorter than the time interval between two noisy operations. $\rho_A$ is the initial state of the actuator, which could be chosen as $\openone_A/2$. When the unitary operation corresponding to $\mathcal{U}'$ is $U=\openone$, only $\mathcal{I}'$ will be performed on actuator qubits. If $H_Q$ is the effective Hamiltonian, and the desired operation $R=e^{-iH_Qt/\hbar}$, the number of noisy operations $N$ is the largest integer that does not exceed $tf$. It is similar for decoupling operations, where $\mathcal{U}'\mathcal{I}'$ is replaced by the twirling operation $\mathcal{E}_1$.

In the ideal case, i.e.~the frequency of noisy operations is infinitely high, the operation $\mathcal{R}' = \mathcal{R}$, where the ideal operation $\mathcal{R}(\cdot) = R\cdot R^\dag$. When the frequency is finite, the actually performed operation is different from the ideal operation and can always be expressed as $\mathcal{R}' = \mathcal{E}_R\mathcal{R}$. Here, the superoperator $\mathcal{E}_R$ denotes the errors in the NOCO $R$, which is found using the Choi-Jamiolkowski isomorphism in our numerical simulations. We would like to remark that the error rates of a single-qubit phase gate shown in Fig.~\ref{FidErr} are obtained using the method described in this section.

\subsection{Distillation and error correction}

The high-fidelity initialization and measurement of data qubits could be achieved by repeating the distillation circuit (blue box in Fig.~\ref{CScircuit}). The initialized state and measurement outcome of the data qubit are read from measurement outcomes of the ancillary qubit. If the distillation circuit is repeated for $n$ times in the measurement (initialization) circuit, there are $n$ measurements of the ancillary qubit and a total of $2^n$ different sets of outcomes. Corresponding to each set of $n$ ancillary-qubit measurement outcomes, $\bar{o}$, the input (output) state of the data qubit is $\ket{0}$ with the probability $q_{\bar{o}}$ and $\ket{1}$ with the probability $1-q_{\bar{o}}$. Then, when the outcomes $\bar{o}$ occur, the input (output) state is likely to be $\ket{0}$ if $q_{\bar{o}}>1-q_{\bar{o}}$, or $\ket{1}$ if $q_{\bar{o}}<1-q_{\bar{o}}$. Therefore, the distillation fails with the probability $p_f = \sum_{\bar{o}} p_{\bar{o}}\min\{q_{\bar{o}},1-q_{\bar{o}}\}$, where $p_{\bar{o}}$ is the probability of outcomes $\bar{o}$.

In a FTQC algorithm of the topology-protected cluster state~\cite{RaussendorfAP}, most of cluster-state qubits (vacuum) are measured in the $\sigma^x$ basis, i.e.~proceeded following the circuit shown in Fig.~\ref{CScircuit}. Except for vacuum qubits, some cluster-state qubits (defects) are measured in the $\sigma^z$ basis for defining the computation algorithm, and some other cluster-state qubits (singular qubits) are measured in the basis of $(\sigma^x\pm\sigma^y)/\sqrt{2}$ for inputting magic states. The circuits for defect qubits and singular qubits are slight different from Fig.~\ref{CScircuit}, i.e.~one more single-qubit gate on the data qubit needs to be added after the second $R_X$ gate. Although errors on defect qubits and singular qubits are different from errors on vacuum qubits due to the additional gate, the different is small and does not affect the threshold~\cite{RaussendorfAP}. Therefore, the threshold is mainly determined by errors generated in the circuit in Fig.~\ref{CScircuit}.

By numerically simulating the errors in the circuit in Fig.~\ref{CScircuit} and comparing the effective phase error rate, contributed by both phase errors on the cluster state and errors of effective data-qubit measurements in the $\sigma^x$ basis, with the threshold $3\%$, we find the FTQC thresholds of NOQC shown in Fig.~\ref{threshold}. We would like to remark that, in our model, phase errors on the same sub-lattice~\cite{RaussendorfAP} are almost independent, i.e.~the correlated errors occur with a rate $< 0.03\% \ll 3\%$ near the threshold, hence the correlations only affect the threshold slightly and could be neglected.

\section{Necessary conditions for noises}
\label{AppNC}

In the main text, we have only discussed depolarizing noise in detail. Here we will show that NOQC is not restricted to depolarizing noise. We take the NOQC network shown in Fig.~\ref{network}~(a) as an example. And we suppose that the noise in initialization operations on triangle actuator qubits is $\mathcal{E}_\text{I}$, the noise in Hadamard gates on triangle actuator qubits is $\mathcal{E}_\text{H}$, and the noise in initialization operations on square actuator qubits is $\mathcal{E}_\text{S}$. Then, with frequent initialization operations, a triangle actuator qubit is frozen in the state $\rho_\text{I} = \mathcal{E}_\text{I}(\ketbra{0}{0})$; with frequent combined (initialization + Hadamard gate) operations, a triangle actuator qubit is frozen in the state $\rho_\text{H} = \mathcal{E}_\text{H}(H\rho_\text{I}H)$; and with frequent initialization operations, a square actuator qubit is frozen in the state $\rho_\text{S} = \mathcal{E}_\text{S}(\ketbra{0}{0})$. To implement universal single-qubit gates on register qubits, states $\rho_\text{I}$ and $\rho_\text{H}$ must be polarized in different directions, i.e.~
\begin{equation}
\Tr(\boldsymbol{\sigma} \rho_\text{I}) \times \Tr(\boldsymbol{\sigma} \rho_\text{H}) \neq 0,
\label{C1}
\end{equation}
where
\begin{equation}
\boldsymbol{\sigma} = \sigma_x \mathbf{i} + \sigma_y \mathbf{j} + \sigma_z \mathbf{k}.
\end{equation}
To implement the two-qubit phase gate, we need
\begin{equation}
\Tr(\sigma_z \rho_\text{S}) \neq 0.
\label{C2}
\end{equation}
Equations~(\ref{C1})~and~(\ref{C2}) are necessary conditions for the noise in operations. These conditions are not satisfied only in extreme cases, i.e.~i) a Hadamard gate cannot ever change the polarization of a qubit initialized in the computational basis, or ii) after an initialization operation, states $\ket{0}$ and $\ket{1}$ still occur with the same probability.

\end{document}